# Stationary one-dimensional dispersive shock waves

Yaroslav V. Kartashov[1] and Anatoly M. Kamchatnov[2]

[1]ICFO-Institut de Ciencies Fotoniques, and Universitat Politecnica de Catalunya, Mediterranean Technology Park, 08860 Castelldefels (Barcelona), Spain
[2]Institute of Spectroscopy, Russian Academy of Sciences, Troitsk, Moscow Region, 142190, Russia

We address shock waves generated upon the interaction of tilted plane waves with negative refractive index defect in defocusing media with linear gain and two-photon absorption. We found that in contrast to conservative media where one-dimensional dispersive shock waves usually exist only as nonstationary objects expanding away from defect or generating beam, the competition between gain and two-photon absorption in dissipative medium results in the formation of localized stationary dispersive shock waves, whose transverse extent may considerably exceed that of the refractive index defect. One-dimensional dispersive shock waves are stable if the defect strength does not exceed certain critical value.

Optical shock waves that usually appear upon regularization of sharp wave fronts by the material dispersion (diffraction) were observed in various physical settings. Such waves can be generated not only by localized inputs, as it was done in fibers [1-4] and photorefractive crystals [5], but also upon propagation of dark beams [6] that may develop gradient catastrophe around the intensity deep, accompanied by the emission of shock-like fan of gray solitons [7,8]. Such waves can form even in materials with nonlocal defocusing nonlinearities, although nonlocality severely reduces the expansion velocity of the shock wave [9,10]. One of the well-known approaches for generation of shock waves relies on the introduction of external potentials [11,12]. The interaction of tilted plane waves with such potentials immediately causes the formation of shock waves attached to the potential [11]. Shock wave formation can be found in diverse nonlinear dispersive systems, such as Bose-Einstein condensates with confining trap and repulsive "piston" [13] or moving obstacle [14] external potentials, as well as in gases and fluids. However, in conservative optical and matter-wave systems whose dynamics is governed by the nonlinear Schrödinger equation the stationary oblique shock waves may form only in two transverse dimensions [15,16], while one-dimensional shock waves in such media appear only as essentially nonstationary objects that expand upon evolution.

In this Letter we show that stationary one-dimensional optical shock waves may form upon the interaction of tilted plane waves with refractive index defects imprinted in dissipative media with defocusing nonlinearity, gain, and two-photon absorption. These mechanisms can be found, e.g., in semiconductors, which are commonly used for the fabrication of optical amplifiers, guiding structures, and where two-photon absorption is the dominating mechanism of optical losses [17,18]. It should be stressed that previously stationary one-dimensional dissipative shock waves (undular bores) were obtained only in the context of water wave theory when viscosity is included [19-21].

We model the propagation of laser radiation along the $\xi$ axis of defocusing cubic medium with linear gain, two-photon absorption, and imprinted refractive index defect by the nonlinear Schrödinger equation for the dimensionless light field amplitude $q$:

$$i\frac{\partial q}{\partial \xi} = -\frac{1}{2}\frac{\partial^2 q}{\partial \eta^2} + q|q|^2 - i\sigma_i q|q|^2 + ip_i q + p_r R(\eta)q. \quad (1)$$

Here $\eta, \xi$ are the normalized transverse and longitudinal coordinates, respectively; $\sigma_i$ is the two-photon absorption strength; $p_i$ is the gain parameter; $p_r$ is the defect depth; the function $R(\eta) = \exp(-\Omega \eta^2)$ describes the defect shape. We consider sufficiently broad defects with $p_r > 0$, $\Omega = 0.04$, and suppose that dissipative terms are small, i.e. $p_i, \sigma_i \ll 1$.

In the absence of refractive index defect (at $p_r = 0$) the Eq. (1) admits a solution in the form of a tilted plane wave $q(\eta,\xi) = \chi \exp(i\alpha\eta + ib\xi)$, with the tilt $\alpha$, $b = -\chi^2 - (\alpha^2/2)$, and amplitude $\chi = (p_i/\sigma_i)^{1/2}$ determined by the balance between gain and absorption. The analysis of modulation stability of plane waves with respect to small harmonic perturbations $\sim \exp(ik\eta + \delta\xi)$ with spatial frequency $k$ shows that such plane waves are stable in defocusing media because real parts of eigenvalues of the associated eigenvalue problem $\delta = -p_i \pm [p_i^2 - (k^2/4)(k^2 + 4\chi^2)]^{1/2}$ are negative. One can introduce the critical tilt or "sound velocity" as a positive value of limit $\alpha_{cr} = \lim_{k\to 0}\lim_{p_i,\sigma_i\to 0}(i\delta/k) = \chi$ that corresponds to almost linear part of the dispersion relation $\delta(k)$ at $p_i, \sigma_i \ll 1$, when $p_i/\sigma_i = \chi^2 = \text{const}$. We consider deformations introduced into shapes of tilted plane waves by the negative refractive index defect with $p_r > 0$. The main result of this Letter is that the interaction of plane waves with such defects in dissipative medium may result in the formation of stable one-dimensional dispersive shock waves that do not change shapes upon propagation. Such waves reside on plane wave background, i.e. $q \to \chi \exp(i\alpha\eta + ib\xi)$ at $\eta \to \pm\infty$. However, their shapes in the vicinity of defect depend dramatically on the tilt $\alpha$ (see Figs. 1-3).

When $\alpha < \alpha_{cr}$ (here we set $p_i, \sigma_i = 0.05$, so that the critical tilt $\alpha_{cr} = \chi = 1$), the stationary disturbance profile $q = (w_r + iw_i)\exp(ib\xi)$ (here $w_r, w_i$ are real and imaginary parts of the field) usually has a smooth shape with a very long monotonically decaying left wing attached to the intensity dip located almost completely inside the defect [see Fig. 1(d) showing field modulus distribution $w = (w_r^2 + w_i^2)^{1/2}$ obtained from Eq. (1) using relaxation method]. Such a disturbance can be described analytically as a dispersionless hydrodynamic approximation to Eq. (1). In contrast to conservative case [11], in this approximation the dissipative term $i\sigma_i q|q|^2$ in Eq. (1) now provides additional regularizing effect in hydrodynamic formulation and just this effect prevents formation of "upstream dispersive shocks" for subcritical values of $\alpha$. Then, the left wing at $\eta \to -\infty$ behaves asymptotically as $w = \chi + a\exp(\kappa\eta)$, where

$$\kappa = \frac{2\sigma_i \chi^2 \alpha}{\chi^2 - \alpha^2} \qquad (2)$$

and the disturbance amplitude $a$ depends on the defect depth $p_r$. One can characterize such waves by the dependence of maximal $w_u = \max(w)$ and minimal $w_l = \min(w)$ amplitudes on defect strength $p_r$. Taking into account that $w_{n \to \pm\infty} \to \chi$ it is convenient to introduce also renormalized energy flow $U_r = \int_{-\infty}^{\infty} [w(\eta) - \chi]^2 d\eta$ of shock wave. Notice, that the propagation constant of shock wave is given by $b = -\chi^2 - (\alpha^2/2)$. The wave's amplitude $w_u - w_l$ and renormalized energy flow monotonically increase with increase of $p_r$ [Figs. 1(a) and 1(b)]. For certain $p_r = p_r^{cr}$ the tangential line to $U_r(p_r)$ dependence becomes vertical indicating on the existence of critical defect depth beyond which this wave pattern family ceases to exist (the propagation at $p_r > p_r^{cr}$ is usually accompanied by considerable radiation from defect region without formation of a stationary structure). The critical defect depth monotonically decreases with $\alpha$ and diverges when $\alpha \to 0$ [Fig. 1(c)]. However, at $0.8 \le \alpha \le \alpha_{cr}$ we were unable to reach critical defect depth and obtained stationary disturbances even for strong defects with $p_r > 1$. In this range of $\alpha$ values the disturbance develops feeble oscillations on its right wing [Fig. 2(a)]. Notice that while at $\alpha \to 0$ the amplitude of disturbance $w_u - \chi$ on the left wing of shock wave is small, at $\alpha \to \alpha_{cr}$ it becomes comparable with the amplitude of oscillation $\chi - w_l$ on the right wing [Fig. 2(a)]. Besides that, the steepness of the disturbance increases with growth of $\alpha$ in the subcritical regime $(0 < \alpha < \alpha_{cr})$, so that Eq. (2) loses its applicability in the limit $\alpha \to \alpha_{cr}$. This means that we cannot neglect diffraction in the description of such a sharp profile. As is known, diffraction effects lead to development of fringes in the intensity profile so that the character of the disturbance profile drastically changes at $\alpha > \alpha_{cr}$.

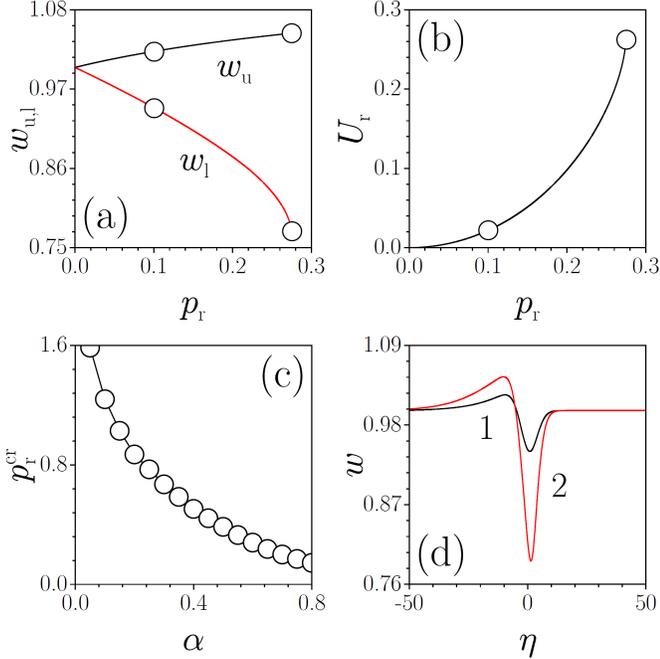

Fig. 1. Maximal and minimal amplitudes of shock wave (a) and renormalized energy flow (b) versus $p_r$ at $\alpha = 0.6$. (c) Critical defect depth for shock wave existence versus $\alpha$. (d) Profiles of stationary shock waves at $p_r = 0.10$ (curve 1) and $p_r = 0.27$ (curve 2) corresponding to $\alpha = 0.6$ and circles in (a) and (b). Left wings decay exponentially according to Eq. (2) where $\kappa$ does not depend on $p_r$; its theoretical value $\kappa_{th} = 0.0938$ agrees rather well with numerical result $\kappa_{num} = 0.0974$.

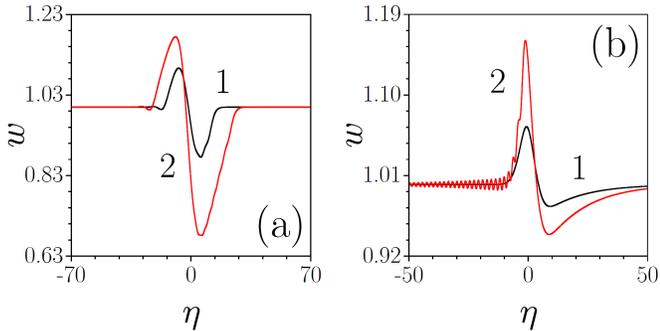

Fig. 2. Profiles of stationary shock waves corresponding to (a) $\alpha = 1$ for $p_r = 0.2$ (curve 1) and $p_r = 0.6$ (curve 2) and to (b) $\alpha = 2$ for $p_r = 0.4$ (curve 1) and $p_r = 0.7$ (curve 2).

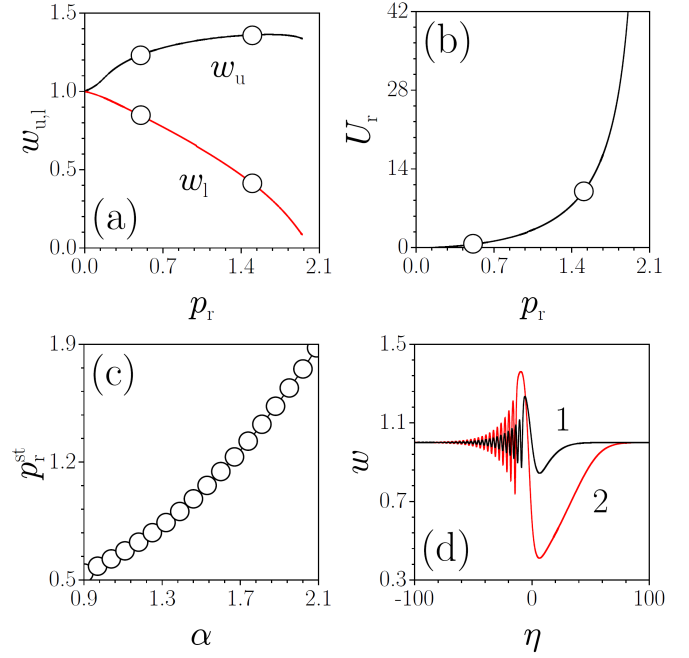

Fig. 3. Maximal and minimal amplitudes of shock wave (a) and renormalized energy flow (b) versus $p_r$ at $\alpha = 1.4$. (c) Threshold defect depth for shock wave stability versus $\alpha$. (d) Profiles of shock waves at $p_r = 0.5$ (curve 1) and $p_r = 1.5$ (curve 2) corresponding to $\alpha = 1.4$ and circles in (a) and (b). Asymptotic theoretical period of oscillations (3) on left wing $d_{th} = 3.206$ agrees well with numerical result $d_{num} = 3.226$. Theoretical $\kappa_{th} = 0.0729$ and numerically calculated $\kappa_{num} = 0.0738$ decay rates of envelopes are also close.

In the case $\alpha > \alpha_{cr}$ one observes the formation of dispersive shock waves with extended oscillations on the left wing [see Fig. 3(d) for typical examples]. Remarkably, such oscillations arising due to regularizing action of diffraction for supercritical light "flows" exponentially decay at $\eta \to -\infty$ and do not expand away from the defect upon propagation along the $\xi$ axis (i.e., shock waves obtained here are fully stationary), as it would occur in conservative systems without dissipation and gain [11]. This oscillatory structure can be described in the framework of the Whitham theory for perturbed integrable equations [22]. In particular, it predicts asymptotic (at $\eta \to -\infty$) wave form $w = \chi + a\exp(\kappa\eta)\cos(2\pi\eta/d)$, where period and decay rate of oscillations are given by:

$$d = \frac{\pi}{(\alpha^2 - \chi^2)^{1/2}}, \quad \kappa = \frac{\sigma_i \chi^2 \alpha}{\alpha^2 - \chi^2}. \qquad (3)$$

The oscillating left wing may expand far beyond the defect region and it becomes more pronounced with the increase of defect depth and for larger tilts $\alpha$. The total amplitude $w_u - w_l$ of shock wave monotonically grows with $p_r$ [Fig. 3(a)]. In contrast to the case $\alpha < \alpha_{cr}$ the field modulus in the deep can decrease almost to zero for high $p_r$ values indicating on the dramatic deformation of the plane wave caused by the defect. The right wing of shock wave at $\eta \to +\infty$ can again be described by Eq. (2) and it can be much wider than the defect. The renormalized energy flow rapidly grows with increase of $p_r$ and shows the tendency for divergence [Fig. 3(b)], but the accurate calculation of critical defect depth at which $U_r \to \infty$ is complicated due to dramatic expansion of the wave with increase of $p_r$ at $\alpha > \alpha_{cr}$. Notice that at small and moderate defect depths the amplitude of oscillations on the left wing of the wave $w_u - \chi$ becomes larger with increase of $\alpha$ than amplitude of oscillations on its right wing $\chi - w_l$ [Fig. 2(b)]. Importantly, one-dimensional shock waves may be stable. The stability was studied by substitution of the perturbed shock-wave solution $q = [w_r + iw_i + u\exp(\delta\xi) + iv\exp(\delta\xi)]\exp(ib\xi)$, where $u,v \ll w_r, w_i$ are real and imaginary parts of perturbation, into Eq. (1), its linearization and numerical calculation of perturbation growth rates $\delta$ from the resulting linear eigenvalue problem using standard eigenvalue solver. While at $\alpha < 0.9$ the shock waves are stable for any $p_r$ value, for $\alpha \geq 0.9$ they become unstable if the defect depth exceeds a certain threshold value $p_r^{st}$. This value monotonically increases with $\alpha$ [Fig. 3(c)]. Notice that increasing the strength of dissipative terms in Eq. (1) results in the reduction of the amplitude of shock waves and faster decay of their tails (thus, at $p_r = 0.2$, $\alpha = 1$ the wave's amplitude $w_u - w_l \approx 0.22$ at $p_i, \sigma_i = 0.05$, while at $p_i, \sigma_i = 0.2$ one has $w_u - w_l \approx 0.09$).

The results of linear stability analysis were confirmed by the direct propagation of perturbed shock waves. While at $p_r < p_r^{st}$ and $\alpha > \alpha_{cr}$ the shock wave with multiple oscillations exhibit stable propagation (Fig 4, top left panel), at $p_r > p_r^{st}$ the exponentially growing small-scale oscillations develop on the right wing of the wave, leading to its instability (Fig. 4, top right panel). At the same time for $\alpha = 0.6 < \alpha_{cr}$ the wave remains stable even for $p_r$ values close to the upper border of the entire existence domain (Fig. 4, bottom panel).

Summarizing, we predicted the possibility of formation of stationary one-dimensional shock waves in dissipative system with linear gain and two-photon absorption upon interaction of tilted plane wave with refractive index defect.

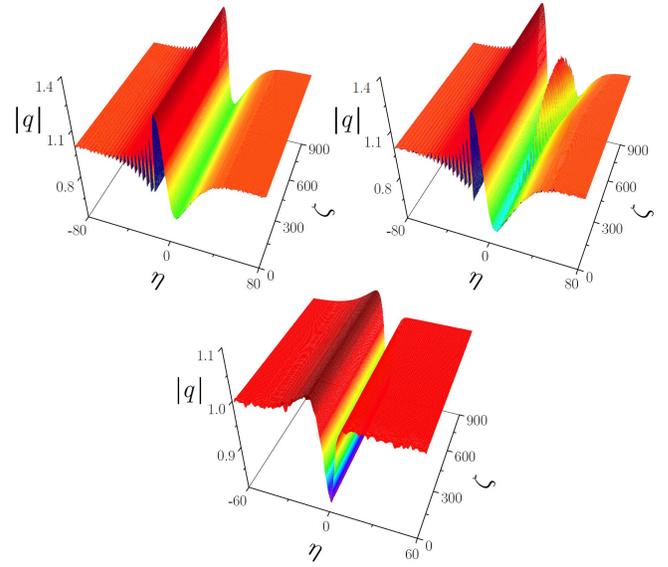

Fig. 4. Dynamics of propagation of perturbed shock waves at $\alpha = 1.4$, $p_r = 0.9$ (top left panel), $\alpha = 1.4$, $p_r = 1.1$ (top right panel), and $\alpha = 0.6$, $p_r = 0.25$ (bottom panel).